\documentclass[prl,aps,twocolumn,superscriptaddress,amsmath,amssymb]{revtex4}
\usepackage{amsmath}
\usepackage[english]{babel}
\usepackage{graphicx}
\usepackage{bm}

\newcommand{\ds}{\displaystyle}

\begin{document}

\title{Temperature controlled FFLO instability in superconductor-ferromagnet hybrids}

\author{S. V. Mironov}
\affiliation{Institute for Physics of Microstructures, Russian Academy
of Sciences, 603950 Nizhny Novgorod, GSP-105, Russia}
\author{D. Yu. Vodolazov}
\affiliation{Institute for Physics of Microstructures, Russian Academy
of Sciences, 603950 Nizhny Novgorod, GSP-105, Russia}
\author{Y. Yerin}
\affiliation{Institute for Physics of Microstructures, Russian Academy
of Sciences, 603950 Nizhny Novgorod, GSP-105, Russia}
\affiliation{Physics Division, School of Science and Technology, Università di Camerino Via Madonna delle Carceri 9, I-62032 Camerino (MC), Italy}
\author{A. V. Samokhvalov}
\affiliation{Institute for Physics of Microstructures, Russian Academy
of Sciences, 603950 Nizhny Novgorod, GSP-105, Russia}
\author{A. S. Mel'nikov}
\affiliation{Institute for Physics of Microstructures, Russian Academy
of Sciences, 603950 Nizhny Novgorod, GSP-105, Russia}
  \affiliation{Lobachevsky State University of Nizhny Novgorod, 23 Prospekt Gagarina, 603950, Nizhny Novgorod, Russia}
\author{A. Buzdin}
\affiliation{University Bordeaux, LOMA UMR-CNRS 5798, F-33405 Talence Cedex, France}
\affiliation{Department of Materials Science and Metallurgy, University of Cambridge, CB3 0FS, Cambridge, United Kingdom}
\affiliation{Sechenov First Moscow State Medical University, Moscow, 119991, Russia}

\begin{abstract}
We show that a wide class of layered superconductor-ferromagnet (S/F) hybrids demonstrate the emergence of the Fulde-Ferrell-Larkin-Ovchinnikov (FFLO) phase well below the superconducting transition temperature. Decreasing the temperature one can switch the system from uniform to the FFLO state which is accompanied by the damping of the diamagnetic Meissner response down to zero and also by the sign change in the curvature of the current-velocity dependence. Our estimates show that an additional layer of the normal metal (N) covering the ferromagnet substantially soften the conditions required for the predicted FFLO instability and for existing S/F/N systems the temperature of the transition into the FFLO phase can reach several Kelvins.
\end{abstract}

\maketitle

In 1964 Fulde, Ferrell, Larkin and Ovchinnikov theoretically showed that strong magnetic field acting on the electron spins in low-dimensional superconductors induces a peculiar non-uniform superconducting phase with the spatial modulation of the order parameter (FFLO phase) \cite{FF, LO}. The key ingredient for the FFLO state formation is the splitting of the Fermi surfaces for the spin-up and spin-down electrons due to the Zeeman interaction. In this case the Cooper pair cannot be constructed from the electrons with the opposite momenta any more, and the total momentum of the pair becomes nonzero. The resulting non-uniform profile of the superconducting gap strongly depends on the sample dimensionality and the anisotropy of the superconductor \cite{MatsudaReview}.

In spite of the transparent physics behind the FFLO instability its experimental observation appeared to be extremely challenging. First, one needs to deal with the low-dimensional samples or with the layered heavy-fermion compounds in order to damp the orbital effect which usually dominates over the Zeeman interaction and suppresses the superconductivity at the magnetic fields well below the FFLO instability threshold \cite{Gruenberg, Shimahara}. Second, the FFLO phase is known to be very sensitive to the disorder which is typically rather strong in thin films or layered superconductors \cite{Aslamazov, Takada}. As a result, the convincing evidence of the FFLO states formation in an external magnetic field has been provided only for some quasi-two-dimensional organic superconductors such as $\lambda{\rm -(BETS)_2GaCl_4}$ \cite{Coniglio}, $\lambda{\rm -(BETS)_2FeCl_4}$ \cite{Uji}, $\kappa{\rm -(BEDT-TTF)_2Cu(NCS)_2}$ \cite{Lortz, Bergk, Wright, Mayaffre, Agosta}, and $\beta^{\prime\prime}{\rm -(ET)_2SF_5CH_2CF_2SO_3}$ \cite{Beyer, Koutroulakis}. The layered structure of these compounds damps the orbital effect for the field orienation parallel to the layers, while the highly anisotropic Fermi surface is expected to provide an additional stability for the FFLO phase \cite{MatsudaReview}.

Another promising possibility to realize the FFLO pairing appears in the multilayered superconductor-ferromagnet (S/F) structures where the interfaces between the layeres are transparent for the electrons \cite{Buzdin_RMP}. In such sandwiches the splitting of the Fermi surfaces occurs due to the exchange field inside the F-layer which does not produce the orbital currents. As a result, the Cooper pair wave function becomes modulated across the layers and the FFLO phase appears. This leads to a number of unusual phenomena such as the oscillatory dependence of the critical temperature of the S/F bilayer on the F-layer thickness \cite{Jiang, Zdravkov} or the $\pi$-junctions formation \cite{Buzdin_Pi, Ryazanov}. The rich interference physics coming from the interplay between the FFLO oscillations period and the layers thicknesses as well as the unusual spin patterns arising in such systems make them attractive for the superconducting spintronics \cite{LinderRev, EschrigRev}.

For more than two decades it was believed that in S/F sandwiches the Cooper pair wave function is always modulated only in the direction  perpendicular to the layers due to the in-plane system homogeneity. But recently it was demonstrated that the spin-triplet superconducting correlations emerging in such system favors the formation of the in-plane FFLO phase with the gap potential modulated along the layers \cite{Mironov}. As a result, the critical temperature for the FFLO phase in the certain range of parameters becomes higher than the one for the uniform state, and the transition from the normal to the FFLO state occurs. Remarkably, the emergence of the in-plane FFLO phase should reveal itself through the vanishing Meissner response of the sample on the external parallel magnetic field. Experimentally, such feature can be detected, e.g., in the surface inductance measurements on the basis of the two-coil technique \cite{Turneaure_1, Turneaure_2} which has been recently applied for the study of the screening properties of the S/F bilayers \cite{Lemberger1, Lemberger2}. The similar instabilities of the uniform state has been predicted also for a ferromagnetic cylinder covered by the superconducting shell \cite{Samokh1, Samokh2, Samokh3} as well as for the planar superconductor / normal metal (N) structures under non-equilibrium quasiparticle distribution \cite{Bobkova, VolkovNE}. However, it appeared that the experimental observation of the in-plane FFLO states in all these systems is hampered by the rigid restrictions for the required material characteristics.

In this paper we predict the existence of the in-plane FFLO phase well-below the critical temperature in a wide class of the thin-film S/F and S/F/N sandwiches. The phase diagrams of such hybrids demonstrate several very unusual features which, to our knowledge, contrast with the diagrams of the all-known systems supporting the FFLO states. Specifically, the FFLO domain can be totally isolated from the phase transition line between the normal and the uniform superconducting states. Decreasing the temperature one can provoke the phase transition between the uniform and FFLO states which is accompanied by the vanishing Meissner response on the in-plane magnetic field and the sign change in the curvature of the current-velocity dependence. Our estimates show that the conditions required for the predicted FFLO instability are rather soft and can be fulfilled for a large number of the existing S/F/N systems consisting, e.g., of the superconducting NbN, MoN, ${\rm MgB_2}$, NbTi, TaN or WSi layer, the ferromagnetic CuNi, PdFe, FeNi or Gd layer, and the layer of Au, Ag, Al or Cu as a normal-metal. For such systems the critical temperature of the transition into the modulated state can reach several Kelvins which makes them very promising for the experimental observation of the FFLO phase.

We start from the general arguments illustrating the origin of the low-temperature FFLO phase formation. 
 Consider a thin-film S/F sandwich of the thickness much smaller that the London penetration depth $\lambda$.
The condition of the gauge invariance of the free energy functional allows us to establish an equivalence of the sign change of the total magnetic response of the thin-film structure (i.e. the quantity $\lambda^{-2}$ averaged across the structure) and the free energy instability towards the formation of the state with a finite in-plane phase gradient. This general recipe is valid for arbitrary temperatures and is nicely confirmed by further direct numerical calculations of the full free energy.
For temperatures $T$ near the superconducting transition temperature $T_c$ the screening parameter $\lambda^{-2}$ which determines the relation ${\bf j}_s=-(\lambda^{-2}/4\pi){\bf A}$ between the superonducting current ${\bf j}_s$ and the vector potential ${\bf A}$ can be expanded in the small parameter $\tau=(T_c-T)/T_c$:
\begin{equation}\label{LambdaTExpansion}
\lambda^{-2}=\chi \tau +\kappa \tau^2,
\end{equation}
where the coefficients $\chi$ and $\kappa$ are temperature independent. In the absence of the F layer the standard BCS model gives $\chi>0$ and $\kappa<0$. At the same time, the exchange field in the ferromagnet gives rise to the spin-triplet superconducting correlations which renormalize these coefficients. For rather large normal state conductivity of the F-layer and small thickness of the S film the coefficient $\chi$ becomes strongly damped and can even vanish. The latter fact indicates the formation of the in-plane FFLO state at $T=T_c$ \cite{Mironov}. It is important that the coefficient $\kappa$ should remain negative reflecting the decrease in the number of quasiparticles when decreasing the temperature. As a result, even for $\chi>0$ there exists a possibility for vanishing of $\lambda^{-2}$. If the total thickness of the S/F sandwich is much smaller than the London penetration depth the part of the free energy containing the square of the superconducting phase gradient is proportional to the $\lambda^{-2}$ value averaged across the structure. Thus, for $|\kappa| \gg\chi$ in Eq.~(\ref{LambdaTExpansion}) the FFLO phase can emerge at the temperature $T_F$ well below $T_c$: $T_F/T_c=1-\chi/|\kappa|$. It is exactly this FFLO instability which makes impossible to observe the global paramagnetism predicted in \cite{Bergeret_PME, Asano, Yokoyama}. The latter paramagnetic state just does not correspond to the free energy minimum \cite{supp}.

To provide a support for the above qualitative arguments we perform an explicit microscopic calculation of the magnetic screening parameter $\lambda^{-2}$ for the dirty S/F bilayer. Our analysis is based on the non-linear Usadel equation
\begin{equation}\label{UsadelGeneral}
-D\left(g\partial_x^2f-f\partial_x^2g\right)+2(\omega_n+ih)f-2\Delta=0
\end{equation}
with the normalization condition
\begin{equation}\label{UsadelNorm}
g^2+ff^\dag=1.
\end{equation}
Here $g(x,\omega_n,h)$ and $f(x,\omega_n,h)$ are the normal and anomalous Green functions respectively, $f^\dag(x,\omega_n,h)=f^*(x,\omega_n,-h)$, $\Delta(x)$ is the superconducting gap potential in the S-layer, $h$ is the exchange field in the F-layer, $\omega_n=\pi T(2n+1)$ are the Matsubara frequencies, and $D$ is the diffusion coefficient of the corresponding layer. The anomalous function and the gap potential also satisfy the self-consistency equation
\begin{equation}\label{UsadelNorm}
\Delta ~{\rm ln}\frac{T}{T_{c0}}+\sum\limits_{n=0}^{\infty}\left(\frac{\Delta}{n+1/2}-2\pi T ~{\rm Re}f\right)=0,
\end{equation}
where $T_{c0}$ is the critical temperature of the isolated superconducting layer. Assuming the small thickness $d_0=(d_s+d_f)\ll \lambda$ where $d_s$ and $d_f$ are the thicknesses of the S and F layers respectively, we may write the London screening parameter averaged over the structure thickness in the form (see, e.g., \cite{Houzet})
\begin{equation}\label{LambdaDef}
\lambda^{-2}=\frac{16\pi^3 T}{ec\Phi_0 d_0}\sum\limits_{n=0}^{\infty}\int\limits_{-d_s}^{d_f}\sigma~{\rm Re} (f^2)dx,
\end{equation}
where $\sigma$ is the normal-state conductivity which takes the value $\sigma_s$ ($\sigma_f$) for the S (F) layer, $\Phi_0=\pi\hbar c/e$ is the magnetic flux quantum.

\begin{figure}[t!]
\includegraphics[width=0.23\textwidth]{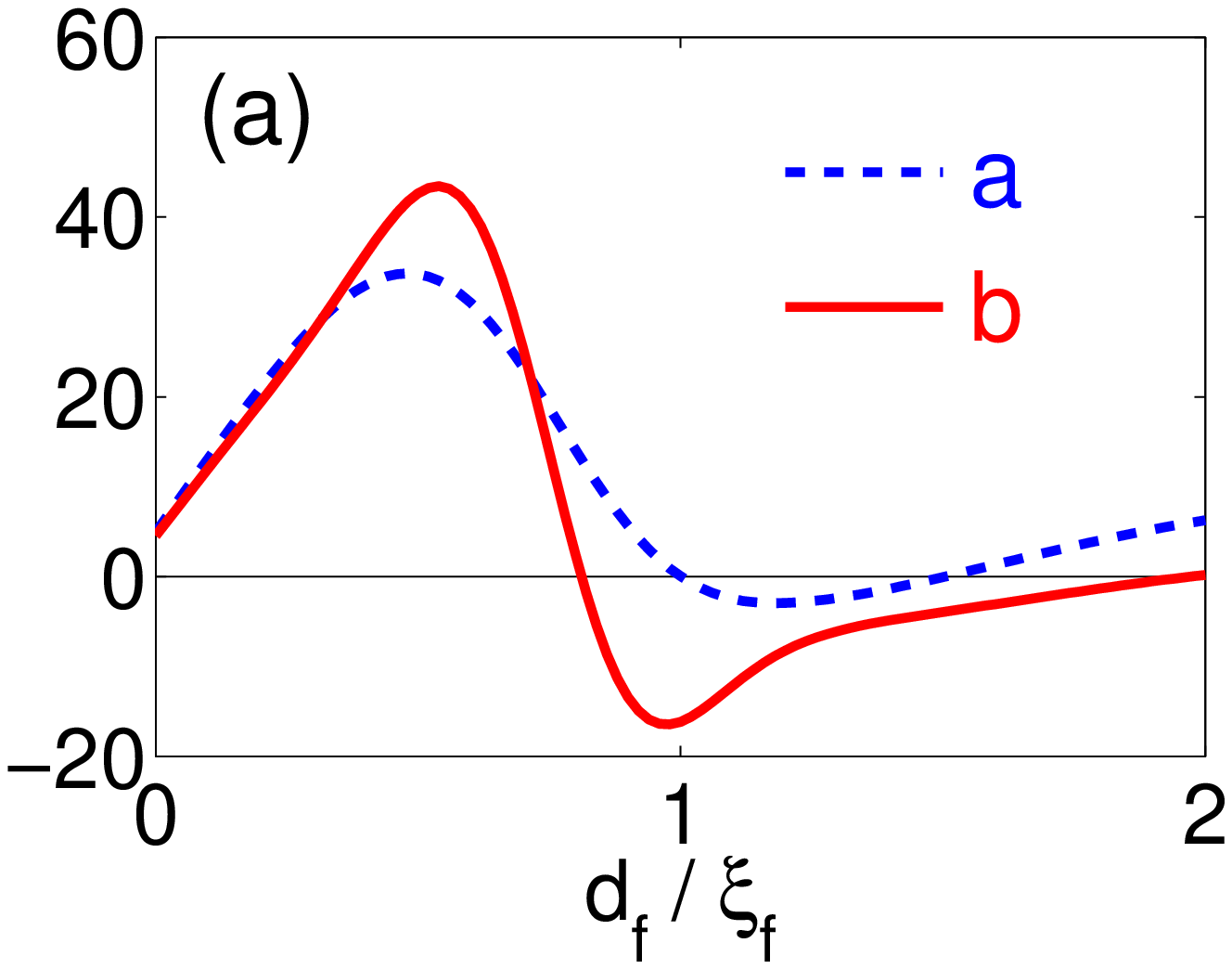}
\includegraphics[width=0.23\textwidth]{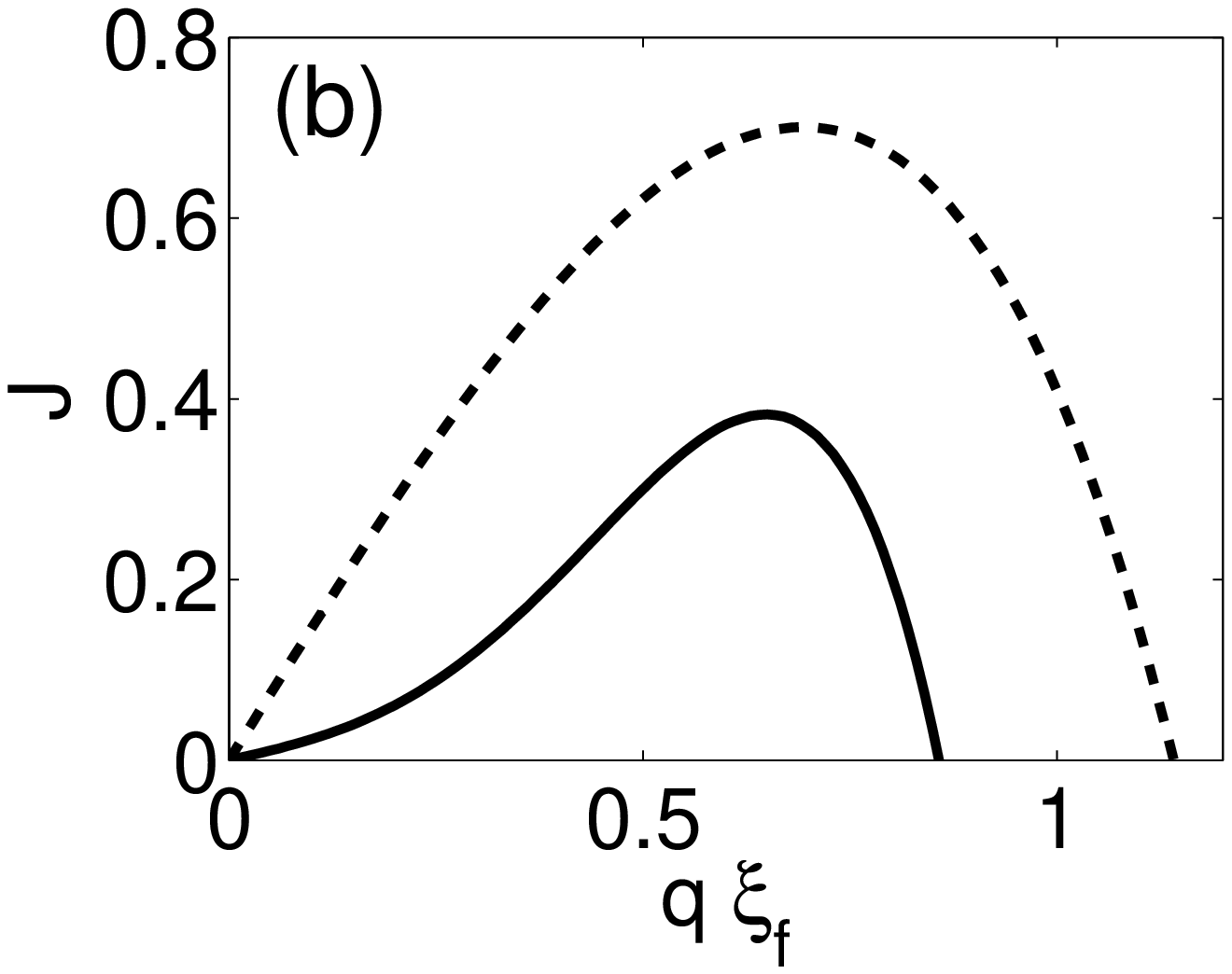}
\caption{(a) The coefficients $a=\alpha(2\pi T_{c})^2\lambda_c^2$ and $b=\beta(2\pi T_{c})^4\lambda_c^2$ in the expansion (\ref{LambdaExpansion}) as  functions of the F-layer thickness. (b) The dependencies of the supercurrent $J=j_s\lambda_c^2 T_c/(\xi_f T_{c0})$ on the superconducting velocity $q$ for $T=0.8T_c(d_f)$, where $\xi_f=\sqrt{D_f/h}$ is the superconducting coherence length inside the ferromagnet and $T_c(d_f)$ is the critical temperature corresponding to the thickness $d_f$ of the F layer. The solid (dashed) line corresponds to $d_f=0.91\xi_f$ ($d_f=0.3\xi_f$).  Here we define $\lambda_c^2=ec\Phi_0d_0/(16\pi^3\sigma_sd_sT_c)$ and take $\xi_f=10\xi_s$, $\sigma_sd_s/(\sigma_f\xi_f)=0.06$. } \label{Fig_Coeff}
\end{figure}

Technically, it is more convenient to rewrite the expansion (\ref{LambdaTExpansion}) in terms of the small temperature-dependent gap potential $\Delta(T)$ \cite{supp}:
\begin{equation}\label{LambdaExpansion}
\lambda^{-2}=\alpha \Delta^2(T)+\beta \Delta^4(T),
\end{equation}
where the coefficients $\alpha$ and $\beta$ do not depend on temperature, $\Delta(T)$ vanishes at $T=T_c$ and monotonically increases with the decreasing $T$.

First, we derive the coefficients $\alpha$ and $\beta$. To do this we assume that the thickness of the superconductor is small: $d_s\ll \xi_s$ where $\xi_s=\sqrt{D_s/(2\pi T_{c0})}$. This allows us to neglect the spatial variations of the gap potential across the S-layer. Also we assume that $h\gg T_{c0}$. For this limit the coordinate dependence of the anomalous Green function has been previously calculated in Ref. \cite{Samokhvalov} up to the terms $\sim O(\Delta^4)$. Substituting this expansion for $f$ into Eq.~(\ref{LambdaDef}) and performing the straightforward calculations we find the analytical expressions for $\alpha$ and $\beta$ which are presented in \cite{supp}. The typical dependencies of these coefficients on the F-layer thickness are shown in Fig.~\ref{Fig_Coeff}(a). If the ratio $\sigma_f/\sigma_s$ is large enough the coefficient $\alpha$ can become negative for certain range of $d_f$ values which signals the formation of the FFLO state at the critical temperature. However, at the points where $\alpha=0$ the coefficient $\beta$ is always negative. As a result, even for small positive $\alpha$ values the second term in Eq.~(\ref{LambdaExpansion}) fully compensate the term $\propto \Delta^2$ at a certain temperature below $T_c$ making $\lambda^{-2}$ vanish and the FFLO phase appear. This finding opens a new perspective for the experimental observation of the transitions between the uniform and FFLO phases since they can be controlled by the variation of temperature.

Another intriguing feature associated with the low-temperature FFLO instability is the sign reversal in the non-linear contribution to the relation between the supercurrent ${\bf j}_s$ and the superconducting velocity which is proportional to the value ${\bf q}=\nabla\varphi-(2\pi/\Phi_0){\bf A}$ ($\varphi$ is the phase of the superconducting order parameter $\Delta$). Qualitatively, this phenomenon can be understood within the Ginzburg-Landau model. Near the transition to the FFLO phase the superconducting contribution to the density of the free energy has the from \cite{Kachkachi}
\begin{equation}\label{GL}
F=\left[-\alpha_0\tau+\beta_0q^2+\delta_0q^4\right]\Delta^2+\left(\gamma_0+\eta_0q^2\right)\Delta^4,
\end{equation}
and $j_s\propto \partial F /\partial q$. At a fixed small $q$ the minimization of the free energy with respect to $\Delta$ gives $\Delta^2=\left[\alpha_0\tau-\beta_0q^2\right]/(2\gamma_0)+O(q^4)$. Substituting this expression into the supercurrent we get: $j_s\propto\left[\beta_0\alpha_0\tau q-\beta_0^2q^3+2\delta_0\alpha_0\tau q^3\right]/\gamma_0$. Far from the FFLO phase domain the last term in the expression for $j_s$ is negligibly small compared to the second one, and $\partial^2j_s/\partial q^2<0$. However, near the FFLO instability the coefficient $\beta_0$ becomes damped and the last term in $j_s$ with $\delta_0>0$ produces the sign reversal in the curvature of the dependence $j_s(q)$ at small $q$.

To calculate the dependence $j_s(q)$ microscopically one has to replace $\omega_n\to\omega_n+Dq^2/2$ in the Usadel equation (\ref{UsadelGeneral}). Generalizing the expressions for the coefficients $\alpha$ and $\beta$ in Eq.~(\ref{LambdaExpansion}) for $q\neq 0$ and taking into account that $j_s=(\Phi_0/2\pi)q\lambda^{-2}(q)$ we obtain the dependencies $j_s(q)$ which are shown in Fig.~\ref{Fig_Coeff}(b). One sees that for $d_f$ values far from the region of the FFLO instability (dashed curve) the dependence $j_s(q)$ has a standard form with the negative second derivative for all $q$ values. However, in the vicinity of the FFLO phase (solid curve) the curvature of the function $j_s(q)$ changes its sign at small $q$ which can be considered as a precursor of the nearby FFLO transition. Experimentally such change in the curvature sign should reveal itself in the third-harmonics electromagnetic response measurements \cite{Pestov, Baryshev, Samokh4}.

\begin{figure}[b!]
\includegraphics[width=0.48\textwidth]{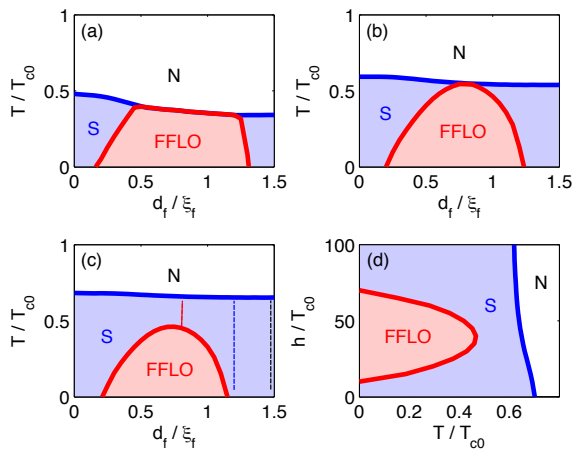}
\caption{(a)-(c) Phase diagrams of the S/F/N sandwiches with $h/T_{c0}=25$ and different thicknesses $d_s$ of the S layer. The ratio $d_s/\xi_0$ with $\xi_0=\sqrt{2\pi}\xi_s$ takes the values: (a) 1.2; (b) 1.4; (c) 1.6. (d) Phase diagram of the S/F/N system with $d_s/\xi_0=1.6$. In all panels $\sigma_f/\sigma_s=1$, $\sigma_n/\sigma_s=150$, $d_n/\xi_0=1$. In the panel (c) the dashed lines indicate the $d_f$ values corresponding to the curves in Fig.~\ref{Fig_SF_Lambda}.} \label{Fig_SF_Diagrams}
\end{figure}

\begin{figure}[t!]
\includegraphics[width=0.235\textwidth]{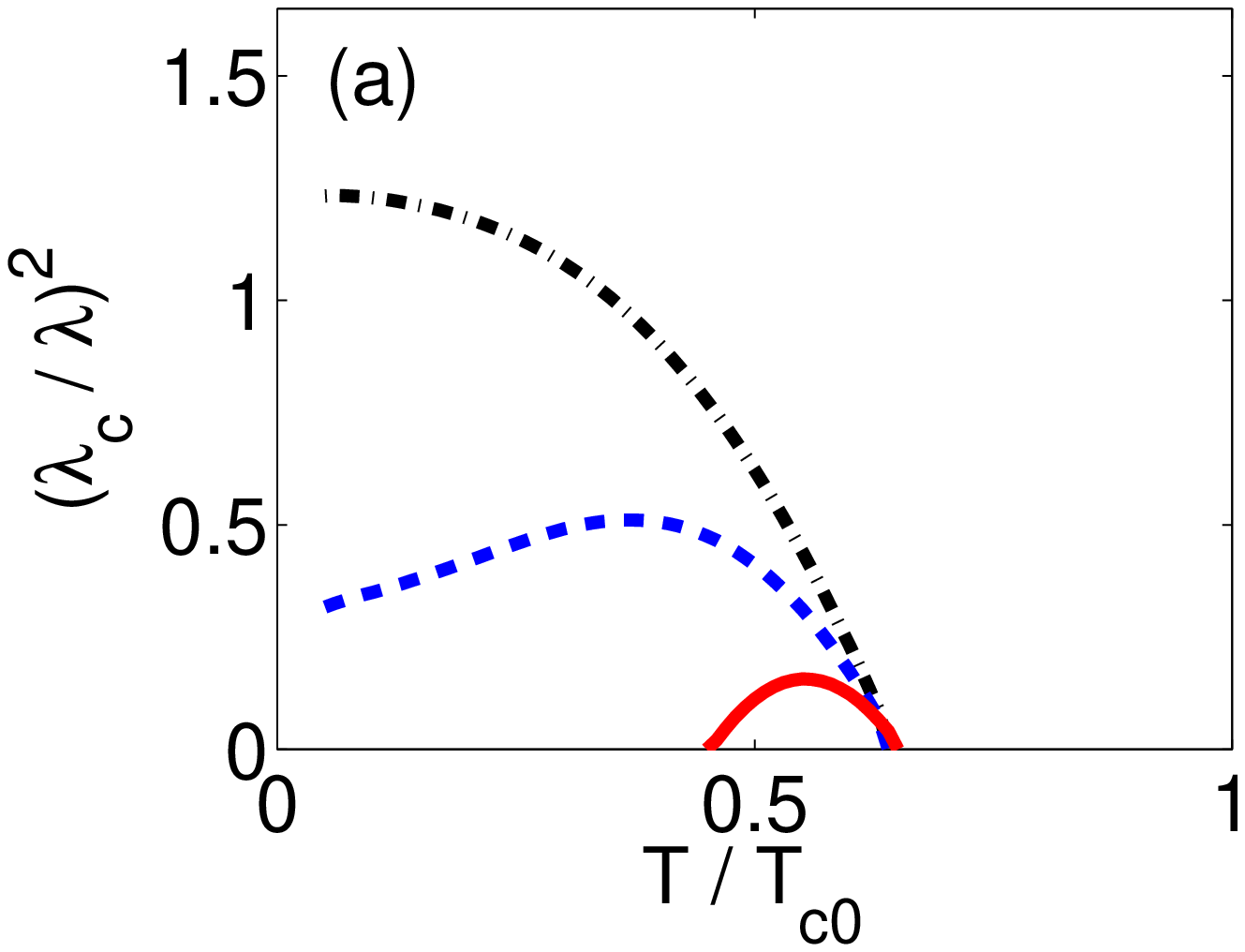}
\includegraphics[width=0.235\textwidth]{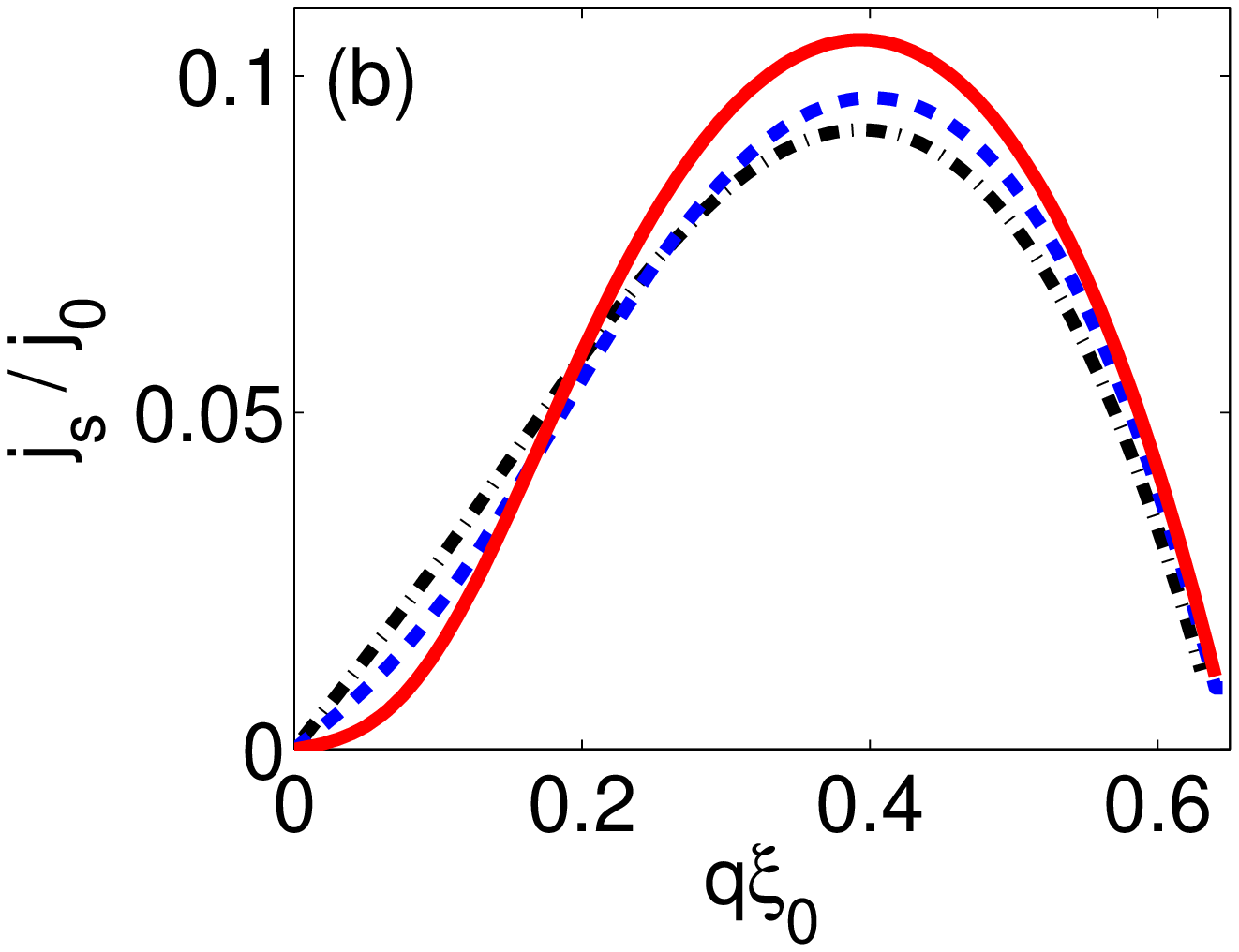}
\caption{(a) The dependencies of the screening parameter $\lambda^{-2}$ on temperature for the S/F/N sandwiches with $d_s/\xi_0=1.6$, the N layer thickness $d_n/\xi_0=1$ and different thicknesses of the F layer. (b) The current flowing along the layers as a function of the superconducting velocity for different $d_f$ values and $T=0.45 T_{c0}$. In both panels the black dash-dotted / blue dashed / red solid curves correspond to $d_f=1.5\xi_f$ / $d_f=1.2\xi_f$ / $d_f=0.8\xi_f$, respectively. The values $\xi_0$ and $j_0$ are defined as $\xi_0=\sqrt{2\pi}\xi_s$ and $j_0=\sigma_s T_c/(e\xi_0)$. Other parameters are the same as in Fig.~\ref{Fig_SF_Diagrams}. } \label{Fig_SF_Lambda}
\end{figure}

All the described analytical results are perfectly supported by the numerical solution of the non-linear Usadel equation \cite{supp} and the direct calculation and comparison of the free energies for the states with different modulation vectors q. The advantage of the numerical approach is its applicability for arbitrary low temperatures in contrast to the above perturbation theory over small $\Delta$ which is limited by the condition $(T_c-T)\ll T_c$. Below we present the numerical results obtained for the S/F/N trilayeres. Previously, in Ref.~\cite{Mironov} it was demonstrated that an additional layer of the normal metal covering the ferromagnet may produce more favorable conditions for the FFLO state formation. The key idea is to choose the thickness of the F-layer to maximize the amplitude of the spin-triplet correlations at the F/N interface. Then if the normal conductivity of the N-layer is large enough the averaged magnetic screening parameter of the sandwich becomes substantially damped favoring the FFLO instability. Here we exploit this idea and show that for a wide class of S, F and N compounds which are typically used in fabrication of the $\pi$-junctions or the spin valves the emergence of the FFLO phase in the S/F/N geometry occurs at the experimentally achievable parameters.

Note also that the transition to FFLO state can be accompanied by the appearance of the in-plane local current density which should average to zero after the integration across the sandwich. The corresponding spontaneous magnetic fields can become of the order of the first critical field $H_{c1}$ which is, of course, a quite measurable value for a variety of experimental techniques. Certainly, these spontaneous currents appear only for a particular profile of the gap function $\Delta=\Delta_0e^{iqy}$ and further studies are necessary to clarify if this state is more energetically favorable than the sinusoidal-like gap profiles.

Note that in our calculations we neglect the contribution of the magnetic field energy which is for sure a valid approximation when the London penetration depth well exceeds the structure thickness.

Fig.~\ref{Fig_SF_Diagrams} shows the series of the phase diagrams of the S/F/N trilayers for different thicknesses $d_s$ of the S-layer. The parts of the red curves below $T_c$ indicate the lines of type-I phase transition between the uniform and the FFLO phases. The increase of $d_s$ results in the shrinkage of the FFLO domain and above the certain threshold this domain can even become fully isolated from the normal state by the region corresponding to the uniform phase. The absence of the boundary between the FFLO and the normal states regions on the $h-T$ phase diagram [see Fig.~\ref{Fig_SF_Diagrams}(d)] contrasts (at least to our knowledge) with the phase diagrams of all previously-known systems  where the direct transition between the normal/FFLO phases can occur.

Finally, Fig.~\ref{Fig_SF_Lambda}(a) demonstrates the typical dependencies of the magnetic screening parameter $\lambda^{-2}$ on temperature for different thicknesses $d_f$ of the ferromagnetic layer. In all cases $d_f$ is chosen in a way that at $T=T_c$ the uniform superconductivity emerges [see Fig.~\ref{Fig_SF_Diagrams}(c)]. There are three qualitatively different types of $\lambda^{-2}$ behavior as the temperature decreases. The first one (black dash-dotted curve) is a monotonic increase of $\lambda^{-2}$ which realizes for the systems parameters far away from the FFLO domain. In contrast, the second one (red solid curve) demonstrates the temperature-induced FFLO phase formation: when decreasing $T$ from $T_c$ the parameter $\lambda^{-2}$ starts to grow, reaches its maximum and then drops down to zero at the point of the FFLO instability. The third one (blue dashed curve) is realized in the intermediate parameter region. Even if the FFLO state does not emerge at any temperatures the dependence $\lambda^{-2}(T)$ can have a maximum which is very unusual for the conventional superconducting systems and serves as a precursor of the nearby FFLO domain. Also our numerical calculations confirm the sign change in the second derivative of the current--velocity dependence $j_s(q)$ near the FFLO domain [see Fig.~\ref{Fig_SF_Lambda}(b) and compare with Fig.~\ref{Fig_Coeff}(b)].

To sum up, we have demonsrated the in-plane FFLO instability well below the critical temperature for S/F and S/F/N hybrids. Experimentally, such instability can be detected by the vanishing Meissner response of the system or by the sign reversal of the third harmonics in the electromagnetic response measurements. At the same time, even outside the FFLO domain on the phase diagram the vicinity to the FFLO instability threshold leads to the unusual non-monotonic dependence of the magnetic screening parameter $\lambda^{-2}$ on temperature. This feature serves as a precursor of the FFLO phase formation. Remarkably, the emergence of the FFLO states in S/F/N sandwiches should occur at the parameter region which can be easily achieved with the wide-spread materials. The combination of superconducting NbN, TaN or WSi layer of the thickness $\sim 10~{\rm nm}$ and the normal metal such as Ag, Au or Al of the thickness $\sim 20-30~{\rm nm}$ gives the normal conductivity ratio $\sigma_n/\sigma_s\sim 150$ which is perfect for the observation of the FFLO states (see, e.g., Fig.~\ref{Fig_SF_Diagrams}). At the same time the relatively high critical temperature of NbN $T_{c0}\sim 10-15~{\rm K}$  makes us hope that the transition to the FFLO phase may correspond to the temperatures of the order of several Kelvins. As usual, the most suitable ferromagnets are CuNi or PdFe which have relatively large coherence lengths providing a possibility to fabricate the layers with $d_f\sim\xi_f$. Thus, the S/F/N sandwiches seem to provide a perfect platform for the observation of the FFLO superconducting states.

Note finally that the above findings presumably can be used to improve the design of kinetic inductance detectors of electromagnetic radiation. Indeed, changing the temperature near the critical temperature of the FFLO transition (where the Meissner screening effect vanishes) one can get rather strong and rapid changes in the kinetic inductance determined by the effective penetration depth and subsequent increase of the detector sensitivity.

\acknowledgments

This work was supported by Russian Science Foundation under Grant No. 15-12-10020 (DYuV, AVS), the French ANR project "SUPERTRONICS" and OPTOFLUXONICS (AB), EU COST CA16218 Nanocohybri (AB), the French-German ANR project "Fermi-NESt" (AB), Russian Presidential Scholarship SP-3938.2018.5 (SVM), Russian Foundation for Basic Research (ASM) and Foundation for the advancement of theoretical physics "Basis" (ASM, SVM, DYuV).

\renewcommand{\theequation}{S\arabic{equation}}

\section*{Supplementary material for ``Temperature controlled FFLO instability in superconductor-ferromagnet hybrids''}

\setcounter{equation}{0}

\subsection{Calculation of $\lambda^{-2}$ as a function of temperature and the superconducting velocity}

To obtain analytical results for the critical temperature
$T_c^{FFLO}$ of the transition between uniform and FFLO states and analyze the dependence of the Meissner screening parameter $\lambda^{-2}$ on the superconducting velocity we
restrict ourselves to the temperatures close to $T_c$.
  The corresponding solution of the Usadel equation is based on a generalization of the results of the Ref.~[S1] for the case of
 nonzero superfluid velocity ($q\neq 0$).
 We assume that $h\gg T_{c0}$ and $d_s\ll \xi_s$, where $\xi_s=\sqrt{D_s/(2\pi T_{c0})}$. The latter condition allows us to neglect the spatial variation of the gap potential $\Delta$ inside the superconductor. In this case all Green function can be expanded over the powers of $\Delta$ up to $\Delta^3$. With this accuracy the expression for the anomalous Green function inside the F-layer which determines the critical temperature and the Meissner screening parameter $\lambda^{-2}$ takes the form:
\begin{equation}\label{F_NL_res}
f(s)=f_0\cosh(ks)\left[1-f_0^2K(s)\right],
\end{equation}
where $s=x/\xi_f$, $k=\sqrt{2i+Q}$, $Q=q^2\xi_f^2$, $q$ is the wave-vector characterizing the superconducting velocity, and the function $K(s)$ is defined as
\begin{equation}\label{R_def}
K(s)=\frac{i\left[4ks\tanh(ks)+12\cosh^2(ks)-9\right]}{32k^2}.
\end{equation}
The constant $f_0$ reads
\begin{equation}\label{f0_def}
f_0=\frac{F_s}{\cosh w}+\frac{F_s^3K(s_f)}{\cosh^3w},
\end{equation}
where $s_f=d_f/\xi_f$, $w=ks_f$. The value $F_s$ is magnitude of the anomalous Green function inside the superconductors which is determined by the following equation:
\begin{equation}\label{FS_Eq}
\left(\omega_n+\nu\right) F_s=\Delta-\frac{1}{2}\left(\Delta F_s^2+\Lambda_0F_s^3\right).
\end{equation}
Here $\varepsilon=\chi\delta$, $\delta=(\sigma_f/\sigma_s)(\xi_f/d_s)$, $\chi=\xi_s^2/\xi_f^2$, $\nu=\varepsilon\pi T_{c0}k\tanh w+\pi T_{c0}\chi Q$ is the pair-breaking parameter,
\begin{equation}\label{L0_def}
\Lambda_0=\frac{i\pi T_{c0}\varepsilon}{4k}\left[\gamma(1+\gamma^2)-w(1-\gamma^2)^2\right]-\pi T_{c0}\chi Q,
\end{equation}
and $\gamma=\tanh w$.

For the further calculations it is convenient to introduce several dimensionless variables:
\begin{equation}\label{Temp_def}
t=T/T_{c0},~~~t_c=T_c/T_{c0},~~~\tau=1-T/T_c.
\end{equation}

First, let us determine the dependence $\tau(\Delta)$ in the vicinity of the superconducting transition. Linearizing the anomalous Green function over $\Delta$, and substituting it into the self-consistency equation we find the expression which implicitly defines the critical temperature $t_c$
\begin{equation}\label{Tc_res}
{\rm ln}t_c=\Psi_0\left(\frac12\right)-{\rm Re}~\Psi_0\left(\frac12+\frac{\Omega}{t_c}\right),
\end{equation}
where $\Psi$ is the Digamma function and $\Omega=\nu/2\pi T_{c0}$. Then from Eq.~(\ref{FS_Eq}) we find the expansion for $F_s$ over $\Delta$ below the transition temperature:
\begin{equation}\label{Fs_expansion}
F_s=\frac{\Delta}{\omega_n+\nu}-\frac{\Delta^3}{2(\omega_n+\nu)^3}-\frac{\Lambda_0\Delta^3}{2(\omega_n+\nu)^4}+O(\Delta^5).
\end{equation}
Substituting Eq.~(\ref{Fs_expansion}) into the self-consistency equation we get:
\begin{equation}
{\rm ln}t=-{\rm Re}\sum\limits_{n=0}^{\infty}\left[\frac{2\pi T\nu}{\omega_n(\omega_n+\nu)} +\frac{\pi T\Delta^2}{(\omega_n+\nu)^3}+\frac{\pi T\Lambda_0\Delta^2}{(\omega_n+\nu)^4}\right],
\end{equation}
Let us denote $\psi_n=\Psi_n\left(1/2+\Omega/t_c\right)$, where $\Psi_n$ is the Polygamma function of the $n$-th order. Then after the expansion of the result over $\tau$ up to the linear terms we find:
\begin{equation}\label{Tau_Delta_SCE}
\tau= \frac{\Delta^2}{(4\pi T_c)^2}\frac{{\rm Re}\left(\psi_2-\frac{\Lambda_0}{6\pi T_c}\psi_3\right)}{{\rm Re}\left(\frac{\Omega}{t_c}\psi_1\right)-1}.
\end{equation}

The obtained explicit expressions for the anomalous Green function in the S and F layers allow to calculate the magnetic screening parameter:
\begin{equation}\label{Lambda1}
\lambda^{-2}=\lambda_c^{-2}\frac{T}{T_c}{\rm Re}\sum\limits_{n=0}^{\infty}\left[F_s^2+\frac{\sigma_f\xi_f}{\sigma_sd_s}\int\limits_{0}^{s_f}f^2(s)ds\right],
\end{equation}
where $\lambda_c^{-2}=16\pi^3 \sigma_sd_sT_c/(ec\Phi_0 d_0)$. Substituting all above expressions for the Green functions inside the S and F layers into Eq.~(\ref{Lambda1}) and performing long but straightforward calculations we finally obtain:

\begin{widetext}

\begin{equation}\label{Lambda3}
\begin{array}{c}{\ds \frac{\lambda^{-2}}{\lambda_c^{-2}}=
\frac{\Delta^2{\rm Re}(\mu\psi_1)}{(2\pi T_c)^2}+
\frac{\Delta^4}{(2\pi T_c)^4}\left[\frac{{\rm Re}\left(\mu\psi_1+
\frac{\Omega}{t_c}\mu\psi_2\right){\rm Re}\left(\psi_2-\frac{\Lambda_0}{6\pi T_c}\psi_3\right)}{4\left[{\rm Re}\left(\frac{\Omega}{t_c}\psi_1\right)-1\right]}
-\frac{{\rm Re}(\mu\psi_3)}{6}+\frac{{\rm Re}(\mu\Lambda_0\psi_4)}{24(2\pi T_c)}+
{\rm Re}\left(\delta\eta\psi_3\right)\right],}
\end{array}
\end{equation}
where
\begin{equation}
\mu=1+\frac{\delta[w(1-\gamma^2)+\gamma]}{2k},~~~ \eta=\frac{i\left\{[w(1-\gamma^2)+\gamma]
\left[(4w\gamma-9)(1-\gamma^2)+12\right]-
2\left[w(1-\gamma^4)+\gamma(2+\gamma^2)\right]\right\}}{192 k^3}
\end{equation}

\end{widetext}

From the obtained resulting expansion (\ref{Lambda3}) for the magnetic screening parameter one can extract both the dependence $\lambda^{-2}(T)$ for $q=0$ and the dependence $\lambda^{-2}(q)$ near $T_c$.

\subsection{Numerical procedure and choice of parameters}

To obtain the profile of the boundaries between the uniform
superconducting state and the FFLO domain on the phase diagram
well below $T_c$ we solve the nonlinear Usadel equation
numerically using the self-consistent iteration procedure. For the
initial distribution $\Delta(x)=const$ we solve Eq. (1) for
Matsubara frequencies $\omega_n=\pi T(2n+1)$ ranging from $n=0$ up
to $n=200$. At the S/F interface the solutions for the Green
functions are matched with Kupriyanov-Lukichev boundary conditions
[S2] while at the outer boundaries of the sample we demand
the zero derivative of the anomalous Green functions. The
numerical procedure is based on the Newton method combined with
the standard matrix multiplication algorithm. We choose the
trigonometrical parametrization of the Green functions which
automatically accounts the normalization condition: $g=cos
\Theta$, $f=sin \Theta $, $f^\dag=sin \Theta$. At each iteration
we substitute the solution for the function $f(x)$ into the
self-consistency equation (4) and calculate the new profile
$\Delta(x)$ (we assume zero superconducting coupling constant
inside the F and N layers so that $\Delta=0$ in these regions).
The iteration repeats until the relative change in $\Delta(x)$
between two iterations becomes less than $10^{-8}$. We use the
value $\xi_0=\sqrt{D_s/T_{c0}}$ as the unit of all lengths and the
value $T_{c0}$ as the unit of energies. The step grid in S the
layer is chosen to be $\delta x= 0.01 \xi_0$. In the F layer it
varies from $0.001 \xi_0$ to $0.02 \xi_0$ depending on the
exchange field $h$ and in the N layer $\delta x=0.01 \xi_0 \ll
\xi_n=\sqrt{D_n/T_{c0}}$. The temperature $T_c^{FFLO}$
corresponding to the transition between the uniform and in-plane
FFLO phases is determined as a solution of equation
$\lambda^{-2}(T=T_c^{FFLO})=0$ with the finite $\Delta$ in S the
layer. The dependence $\lambda^{-2}(T)$ is calculated on the base
of Eq. (5).

At temperature below $T_c^{FFLO}$ and/or in the case when the net
superconducting current is flowing {\it along} the layers we put
an additional term $Dg(x)f(x)q$ with $q=\nabla \varphi-
(2\pi/\Phi_0){\bf A}$ to the left-hand side of Eq. (1) which takes
into account the nonzero velocity of the superconducting
condensate $v_s\propto q/m$ in that direction.

We assume that the density of states in all three layers is the
same and, thus, the ratio of conductivities is proportional to the
ratio of the corresponding diffusion coefficients (e.g.,
$\sigma_s/\sigma_n=D_s/D_n$). This assumption allows the
substantial simplification since the number of the free parameters
becomes strongly decreased.

In the numerical calculations we use the system parameters which are close to the experimentally achievable ones for the real superconducting, normal metal and ferromagnetic films (see Tables 1,2,3). We assume that the exchange field in the F-layer is of the order of the Curie temperature $T_{Curie}$.

\begin{table*}
\caption{\label{tab:table1}Residual resistivity of typical low-resistance metallic films}
\begin{ruledtabular}
\begin{tabular}{cccccccccc}
 Metal & Ag [S3] & Au [S4] & Al [S3] & Cu [S5] \\
d (nm)& 34 &   34.2   &  39   &    30         \\
$\rho_{res}$ $(\mu \Omega\cdot cm)$& 1.2  & 1.0  & 1.5  & 1.2  \\
\end{tabular}
\end{ruledtabular}
\end{table*}

\begin{table*}
\caption{\label{tab:table2} Resistivity and critical temperature of typical dirty superconducting films}
\begin{ruledtabular}
\begin{tabular}{cccccccccc}
Superconductor & NbN [S6] & TaN [S6] & WSi [S7] & MgB$_2$[S8] \\
d (nm)& 10 &   9.7   & 5   &    10        \\
$\rho (T \gtrsim T_c)$ $(\mu \Omega\cdot cm)$& 85  & 108  & 208  & 300  \\
D (cm$^2$/s) & 0.78  & 0.77 & 0.7  &   \\
$T_{c0}$ (K) & 15.2  & 10.7  & 3.88  & 18  \\
$\sqrt{2\pi} \xi_s$ (nm) & 6.2  & 7.4  & 11.7  &   \\
\end{tabular}
\end{ruledtabular}
\end{table*}

\begin{table*}
\caption{\label{tab:table3} Resistivity and exchange field of typical dirty ferromagnetic films}
\begin{ruledtabular}
\begin{tabular}{cccccccccc}
Ferromagnet & Cu$_{0.47}$Ni$_{0.53}$ [S9] & Pd$_{0.84}$Ni$_{0.16}$ [S10] & Ni [S11]   \\
d (nm)& 9-24 &   6-40   & 5          \\
$\rho$  $(\mu \Omega\cdot cm)$& 62  & 43-20  & 190    \\
$T_{Curie}$ (K) & 60  & 190  & 600    \\
\end{tabular}
\end{ruledtabular}
\end{table*}

Our calculations show that the results for the temperature of the FFLO state emergence $T_c^{FFLO}$ is not very sensitive to the ratio between the conductivities of S and F layers [see Fig. 1(a)]. However the conductivity of the normal metal has a strong influence on $T_c^{FFLO}$. When $\sigma_n/\sigma_s$ decreases the area corresponding to the FFLO state on the phase diagram shrinks rapidly [see Fig. 1(b)] and finally the FFLO domain disappears. For example, choosing the parameters $d_s=1.6 \xi_0$, $d_n=\xi_0$, $\sigma_f/\sigma_s=1$ and $h=25 T_{c0}$ such disappearance occurs at $\sigma_n/\sigma_s=30$. At the same time, for $d_s=1.2 \xi_0$ the FFLO phase still exists with the maximal $T_c^{FFLO} \simeq 0.2 T_{c0}$ at $d_f=0.09 \xi_0 \simeq 0.45 \xi_f$.

\begin{figure}[t!]
\includegraphics[width=0.53\textwidth]{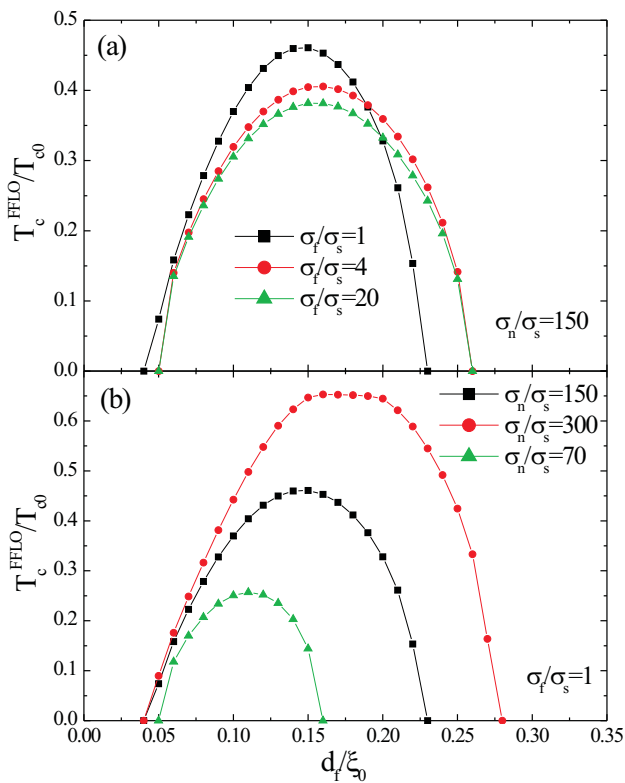}
\caption{Dependence of $T_c^{FFLO}$ on the thickness of the
ferromagnetic layer ($h=25 T_{c0}$): (a) for different values of
$\sigma_f$; (b) for different values of $\sigma_n$. In both cases we choose the thickness
of the S layer $d_s=1.6\xi_0$ and thickness of the N layer $d_n=\xi_0$.}
\end{figure}

We also verify that in the presence of the reasonable barriers between the S, F and N layers the FFLO phase still survives (see Fig. 2). Such barriers are controlled by the dimensionless parameters $\gamma_{SF}=R_{SF}S_{SF}/(\rho_f \xi_0)$ and $\gamma_{FN}=R_{FN}S_{FN}/(\rho_n\xi_0)$ for the S/F and F/N interfaces, respectively. Here $R_{SF}$ and $R_{FN}$ are the electron resistivities of the S/F and F/N boundaries while $S_{SF}$ and $S_{FN}$ are the corresponding boundary areas.

\begin{figure}[hbt!]
\includegraphics[width=0.53\textwidth]{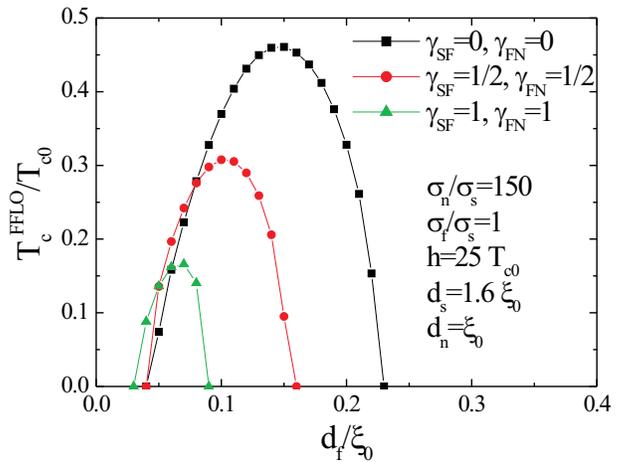}
\caption{The dependence of $T_c^{FFLO}$ on the thickness of the ferromagnetic layer ($h=25 T_{c0}$) for different barriers $\gamma_{SF}$ and $\gamma_{FN}$ between the layers.}
\end{figure}

In Figs. 3(a,b) we demonstrate how $T_c^{FFLO}$ depends on the thicknesses of the N and S layers. The FFLO state may exist in a wide range of $d_s$ and $d_n$ values. The plateau on the dependence $T_c^{FFLO}(d_f)$ in Fig. 3(b) for $d_s=1.2\xi_0$ is connected with the coincidence of $T_c^{FFLO}$ and $T_c$ in this range of $d_f$ values (the same plateau is noticeable in Fig. 1(b) when $\sigma_n/\sigma_s =300$ and in Fig. 2(a) of the main text of the paper).

\begin{figure}[hbt!]
\includegraphics[width=0.53\textwidth]{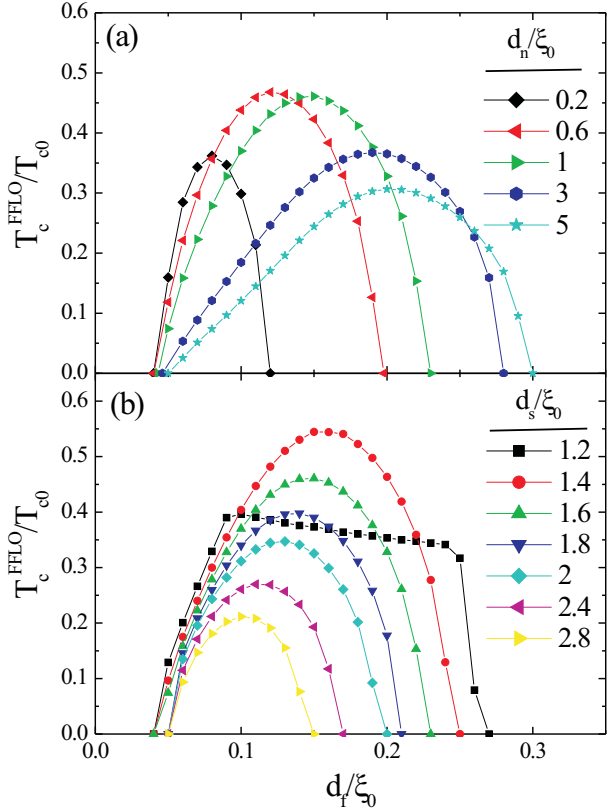}
\caption{The dependence of $T_c^{FFLO}$ on the thickness of the
ferromagnetic layer ($h=25 T_{c0}$): (a) for different $d_n$; (b) for different $d_s$. In both cases $\sigma_n/\sigma_s=150$ and$\sigma_f/\sigma_s=1$.}
\end{figure}

In Fig. 4 we show the dependence of the transition temperature $T_c^{FFLO}$ on the $d_f$ value for the F layers with different exchange fields $h$. The favorable conditions for the FFLO phase observation are realized for the ferromagnets with the small exchange field  $h$ where the corresponding range of the $d_f$ thickness is rather narrow.

\begin{figure}[hbtp]
\includegraphics[width=0.53\textwidth]{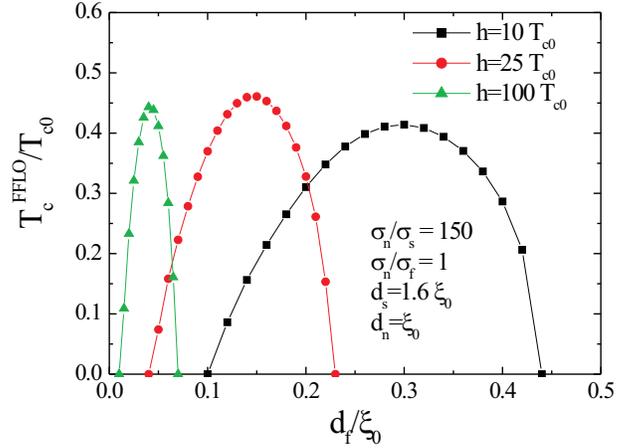}
\caption{The dependence of $T_c^{FFLO}$ on the thickness of the ferromagnetic layer for different values of the exchange field $h$ in the F layer. All material parameters of the S, F and N layers are shown in the figure.}
\end{figure}

\begin{figure}[hbtp]
\includegraphics[width=0.53\textwidth]{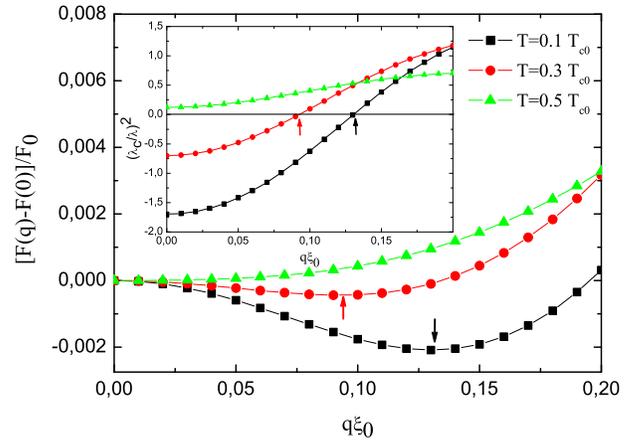}
\caption{ The dependence of the free energy of S/F/N trilayer on the superconducting velocity
for different temperatures. The system parameters are the same as in Fig. 4, $h=25 T_{c0}$ and $d_f=0.8\xi_f$.
In the inset we show the corresponding dependencies of the screening parameter  $\lambda^{-2}$ on the superfluid velocity.}
\end{figure}

And finally, in Fig. 5 we present the excess free
energy  $F$ of S/F/N trilayer in the superconducting state with respect to the one in the normal state as a function of the superfluid velocity  $\propto
q$ for different temperatures. The expression for the excess free energy reads [S12]

\begin{equation}\label{FEnergy}
\begin{array}{c}{\ds
F(q)=\pi N(0)k_BT\sum_{\omega_n \geq 0}\int Re\{\hbar D[(\nabla
\Theta)^2+q^2
sin^2\Theta]}\\{\ds -4(\hbar\omega_n+ih)(cos\Theta-1)-2\Delta sin \Theta
\}dx,}
\end{array}
\end{equation}
where $k_B$ is the Boltzmann constant and $N(0)$ is the density of states at the Fermi level per one spin projection. When performing numerical calculation of the free energy (\ref{FEnergy}) we use trapezoidal rule.

At temperature below $T_c^{FFLO}$ the minimum of the free energy
corresponds to a nonzero $q$ value manifesting, thus,  the
appearance of the FFLO state. Note that in this modulated state
there is a finite magnetic field inside the trilayer (due to the
finite currents flowing in the S and N layers in the opposite
directions with zero average) but its maximal magnitude at the
chosen parameters is of the order of $\Phi_0/2\pi \lambda^2(0)$
and it gives just a small contribution to the free energy $F$
(less than $10^{-4} F_0$, where $F_0 =\pi N(0)(k_BT_c)^2 \xi_0$).
In the inset we present the dependencies $\lambda^{-2}(q)$
[defined by Eq.~(5) of the main text] for different temperatures.
In the FFLO state  $\lambda^{-2}(q)=0$ at all temperatures below
$T_c^{FFLO}$.

\bigskip

\noindent [S1] A. V. Samokhvalov and A. I. Buzdin, Phys. Rev. B \textbf{92}, 054511 (2015).

\noindent [S2] M. Yu. Kupriyanov and V. F. Lukichev, Sov. Phys. JETP \textbf{67}, 1163 (1988).

\noindent [S3] J. W. C. De Vries, Thin Solid Films \textbf{167}, 25 (1988).

\noindent [S4] J. W. C. De Vries, Thin Solid Films \textbf{150}, 201 (1988).

\noindent [S5] Y. P. Timalsina, A. Horning, R. F. Spivey, K. M. Lewis, T.-S. Kuan, G.-C. Wang and T.-M. Lu, Nanotechnology \textbf{26}, 075704 (2015).

\noindent [S6] K. Ilin, D. Henrich, Y. Luck, Y. Liang, M. Siegel, D. Yu. Vodolazov,  Phys. Rev. B \textbf{89}, 184511 (2014).

\noindent [S7] X. Zhang, A. Engel, Q. Wang, A. Schilling, A. Semenov, M. Sidorova,
H.-W. Hubers, I. Charaev, K. Ilin, and M. Siegel, Phys. Rev. B \textbf{94}, 174509 (2016).

\noindent [S8] H. Shibata, H. Takesue, T. Honjo, T. Akazaki, and Y. Tokura, Appl. Phys. Lett. \textbf{97}, 212504 (2010).

\noindent [S9] V. A. Oboznov, V.V. Bol'ginov, A. K. Feofanov, V.V. Ryazanov, and A. I. Buzdin, Phys. Rev. Lett. \textbf{96}, 197003 (2006).

\noindent [S10] C. Cirillo, E. A. Ilyina and C. Attanasio, Supercond. Sci. Technol. \textbf{24}, 024017 (2011).

\noindent [S11] C. J. Kircher, Phys. Rev. B \textbf{168} 437 (1968).

\noindent [S12] M. Eltschka, B. Jack, M. Assig, O. V. Kondrashov, M. A. Skvortsov, M. Etzkorn, C. R. Ast, K. Kern, Appl. Phys. Lett. \textbf{107} 122601 (2015).

\end{document}